# Unsupervised Bayesian Generation of Synthetic CT from CBCT Using Patient-Specific Score-Based Prior

Junbo Peng[1], Yuan Gao[1], Chih-Wei Chang[1], Richard Qiu[1], Tonghe Wang[2], Aparna Kesarwala[1], Kailin Yang[3], Jacob Scott[3], David Yu[1] and Xiaofeng Yang[1*]

[1]Department of Radiation Oncology and Winship Cancer Institute, Emory University, Atlanta, GA 30322

[2]Department of Medical Physics, Memorial Sloan Kettering Cancer Center, New York, NY 10065

[3]Department of Radiation Oncology, Taussig Cancer Center, Cleveland Clinic, Cleveland, OH 44195
Email: xiaofeng.yang@emory.edu

**Running title**: Unsupervised CBCT image correction

# Abstract

**Background:** Cone-beam computed tomography (CBCT) scans, performed fractionally (e.g., daily or weekly), are widely utilized for patient alignment in the image-guided radiotherapy (IGRT) process, thereby making it a potential imaging modality for the implementation of adaptive radiotherapy (ART) protocols. Nonetheless, significant artifacts and incorrect Hounsfield unit (HU) values hinder their application in quantitative tasks such as target and organ segmentations and dose calculation. Therefore, acquiring CT-quality images from the CBCT scans is essential to implement online ART in clinical settings.

**Purpose:** This work aims to develop an unsupervised learning method using the patient-specific diffusion model for CBCT-based synthetic CT (sCT) generation to improve the image quality of CBCT.

**Methods:** The proposed method is in an unsupervised framework that utilizes a patient-specific score-based model as the image prior alongside a customized total variation (TV) regularization to enforce coherence across different transverse slices. The score-based model is unconditionally trained using the same patient's planning CT (pCT) images to characterize the manifold of CT-quality images and capture the unique anatomical information of the specific patient. The efficacy of the proposed method was assessed on images from anatomical sites including head and neck (H&N) cancer, pancreatic cancer, and lung cancer. The performance of the proposed CBCT correction method was evaluated using quantitative metrics including mean absolute error (MAE), peak signal-to-noise ratio (PSNR), and normalized cross-correlation (NCC). Additionally, the proposed algorithm was benchmarked against two other unsupervised diffusion model-based CBCT correction algorithms.

**Results:** The proposed method significantly reduced various kinds of CBCT artifacts in the studies of H&N, pancreatic, and lung cancer patients. In the lung stereotactic body radiation therapy (SBRT) patient study, the MAE, PSNR, and NCC were improved from 91 HU, 27 dB, and 0.98 in the original CBCT images to 50 HU, 31 dB, and 0.99 in the generated synthetic CT (sCT) images. Compared to two other unsupervised diffusion model-based algorithms, the proposed method demonstrated superior performance in artifact reduction.

**Conclusions:** The proposed unsupervised method can generate sCT from CBCT with reduced artifacts and precise HU values, enabling CBCT-guided segmentation and replanning for online ART.

# 1. Introduction

In current image-guided radiotherapy (IGRT) protocols, cone-beam computed tomography (CBCT) scans are routinely performed on a daily or weekly basis for treatment monitoring and patient setup.[1] Compared to diagnostic fan-beam CT images, CBCT images are corrupted by significant artifacts, including streaking, cupping, shading, and scatter contamination, leading to inaccurate Hounsfield unit (HU) measurements.[2-4] Such issues limit the quantitative utilization of CBCT images and impede the practice of CBCT-based adaptive radiotherapy (ART). To overcome this challenge, contemporary ART practices resort to deformed planning CT (dpCT) images aligned with the CBCT anatomy for dose calculations.[5] Nevertheless, the performance of image registration is often contingent upon the operator's expertise and intuition, which in turn constrains the precision of ART applications.

Conventional solutions to CBCT correction aim to reduce artifacts from specific sources via hardware- or algorithm-based scheme. Hardware-based solutions primarily focus on scattering artifacts reduction, such as anti-scatter grid,[6] lattice-shaped beam stopper,[7] and primary-modulation beam filter.[8] Algorithm-based methods attempt to correct artifacts by modeling the physics process of CBCT imaging or by a deep learning model, such as Monte Carlo simulation for scatter correction,[9] and projection- or image-domain neural networks to alleviate scatter,[10,11] metal,[12] and streaking artifacts.[13] Besides the reduced quantum efficiency for hardware modifications and intensive computation for physics modeling,[14,15] these solutions are limited to artifact reduction from specific source, making them less practical in CBCT-based ART implementation.

With the success of deep learning in medical image synthesis, CBCT-based synthetic CT (sCT) generation has been developed as a general solution to CBCT correction. sCT generation ignores the mechanism of different artifacts and directly learns the mapping from CBCT distribution to pCT distribution via neural networks. Generally, the training of neural networks for sCT generation can be categorized as supervised and unsupervised. Supervised methods train the image-to-image translation model using matched CBCT-dpCT image pairs.[16-20] However, the inevitable residual mismatches between CBCT and dpCT can adversely impact the image synthesis performance, making it impossible to create perfectly matched pairs for supervised training. To address this limitation, unsupervised methods have been proposed, which use unpaired CBCT-pCT data for the model training. For example, Chen et al designed an unsupervised CBCT-CT translation network with a customized contextual loss to maintain the anatomical information of CBCT and a style loss to achieve pCT-quality image generation.[21]

In recent years, diffusion models have attracted great attention and shown state-of-the-art (SoTA) performance in the fields of medical imaging and medical image synthesis.[22-26] Denoising diffusion probabilistic model (DDPM),[27] one implementation of diffusion models, has been introduced in CBCT-based sCT generation and shown superior performance than GANs.[28] However, the reported works for sCT generation, including the DDPM-based methods, formulated the CBCT correction task as an implicit image-to-image translation and failed to enforce the data consistency between sCT and CBCT in a mathematically explicit way. Recently, significant progress has been achieved in solving inverse problems using diffusion models.[29-34] These works inspired us that the CBCT correction can be performed in a totally different way that utilizes a diffusion image prior and an explicit data-fidelity objective during the iterative sCT generation.

In this work, we formulated the sCT generation as an inverse problem and proposed a Bayesian framework for iterative CBCT correction. A score-based diffusion model was unsupervised trained to learn the patient-specific prior using pCTs from the same patient. A customized data consistency consisting of CBCT-sCT image fidelity and total-variation (TV) regularization was enforced during the sampling stage of the score-based model, leading to the generation of pCT-quality sCT with the preservation of anatomical information in CBCT. The proposed method was evaluated on CBCT images from H&N, pancreatic, and lung cancer patients, verifying the feasibility of CBCT-based sCT generation. The major contributions of this work include (i) we formulated the sCT generation as an iterative signal reconstruction problem instead of an image-to-image translation task, (ii) we proposed an unsupervised Bayesian framework to solve the problem with a patient-specific image prior, and (iii) we achieved volumetric image generation in a slice-by-slice manner with efficient coherence enforcement between different slices.

## 2. Methodology

### 2.1 Principles of score-based diffusion models

Let $x_0 \sim p_0(x)$ denotes the target data distribution. The forward process of diffusion models progressively perturbs the target sample to a noise sample via Gaussian kernels, and can be formulated by a forward stochastic differential equation (SDE)

$$dx = f(x, t)dt + g(t)dw \tag{1}$$

where $f(x, t)$ is the drift function, $g(t)$ is the scalar diffusion function, and $w$ is the standard Brownian motion. The forward SDE starts from $t = 0$ and obtains a standard Gaussian distribution when $t = T$. For any SDE, there is a reverse SDE in a closed form of

$$dx = \left[f(x, t) - g^2(t)\nabla_{x_t} \log p(x_t)\right]dt + g(t)d\bar{w} \tag{2}$$

which can be used to generate samples in $p_0(x)$ from the Gaussian by reversing the perturbation process from $t = T$ to $t = 0$ via Langevin dynamics. Here $dt$ represents a reverse time step, $\nabla_x \log p_t(x)$ is the (Stein) score function of the distribution $p_t(x)$, and $d\bar{w}$ denotes a standard Brownian motion running backward.

Diffusion models aim to train a parametrized model $s_{\theta,t}(\cdot)$ to estimate the time-dependent score function $\nabla_x \log p(x_t)$,[35] i.e.,

$$s_{\theta,t}(x_t) \approx \nabla_{x_t} \log p(x_t) \tag{3}$$

With the optimized score-based model $s_{\theta,t}(\cdot)$, one can plug it into the reverse SDE in Eq. (2) to simulate the reverse SDE as

$$dx = \left[f(x, t) - g^2(t)s_{\theta,t}(x_t)\right]dt + g(t)d\bar{w} \tag{4}$$

and solve it using numerical solvers, e.g., Euler-Maruyama discretization.

Throughout this work, we adopted the variance-preserving SDE (VP-SDE) to determine the noise schedule in the diffusion process.[36] In VP-SDE, the drift function and scalar diffusion function are parameterized as

$$\begin{cases} \boldsymbol{f}(x,t) = -\dfrac{\beta_t}{2}\boldsymbol{x} \\ g(t) = \sqrt{\beta_t} \end{cases} \tag{5}$$

where $\beta_t$ is the prefixed noise schedule.

## 2.2 Unsupervised sCT generation using score-based diffusion

Theoretically, given a condition $\boldsymbol{y}$, one can obtain the $\boldsymbol{x}_0 \sim p(\boldsymbol{x}_0|\boldsymbol{y})$ by solving the reverse SDE with a conditional score function $\nabla_{\boldsymbol{x}} \log p(\boldsymbol{x}_t|\boldsymbol{y})$. The modified reverse SDE can be written as

$$d\boldsymbol{x} = \left[ -\frac{\beta_t}{2}\boldsymbol{x} - \beta_t \nabla_{\boldsymbol{x}_t} \log p(\boldsymbol{x}_t|\boldsymbol{y}) \right] dt + \sqrt{\beta_t}\, d\bar{\boldsymbol{w}} \tag{6}$$

If precisely paired data are available, e.g., CBCT($\boldsymbol{y}$)-CT($\boldsymbol{x}$) pairs, one can train a conditional score model $\boldsymbol{s}_{\theta,t}(\boldsymbol{x}|\boldsymbol{y})$ to approximate $\nabla_{\boldsymbol{x}} \log p_t(\boldsymbol{x}|\boldsymbol{y})$ via supervised score matching.[35] However, if accurately matched data pairs are unavailable, the conditional score function cannot be directly learned by a diffusion model.

To perform unsupervised sCT generation conditioned on acquired CBCT images, we leveraged the Bayes' rule to split the conditional score function as

$$\nabla_{\boldsymbol{x}_t} \log p(\boldsymbol{x}_t|\boldsymbol{y}) = \nabla_{\boldsymbol{x}_t} \log p(\boldsymbol{x}_t) + \nabla_{\boldsymbol{x}_t} \log p(\boldsymbol{y}|\boldsymbol{x}_t) \tag{7}$$

where the first term is the unconditional score function which can be approximated via unsupervised learning (Eq. 3), and the second term is a data-fidelity term The resultant reverse SDE becomes

$$d\boldsymbol{x} = \left[ -\frac{\beta_t}{2}\boldsymbol{x} - \beta_t \boldsymbol{s}_{\theta,t}(\boldsymbol{x}_t) \right] dt + \sqrt{\beta_t}\, d\bar{\boldsymbol{w}} - \beta_t \nabla_{\boldsymbol{x}_t} \log p(\boldsymbol{y}|\boldsymbol{x}_t)\, dt \tag{8}$$

In the context of CBCT correction, the data consistency is enforced at each time step by minimizing the Euclidean distance between the predicted sCT and acquired CBCT, i.e.,

$$\min_{\widetilde{\boldsymbol{x}}_0} \phi = \|\widetilde{\boldsymbol{x}}_0 - \boldsymbol{y}\|_2^2 + \eta \|\widetilde{\boldsymbol{x}}_0 - \hat{\boldsymbol{x}}_0\|_2^2 \tag{9}$$

where $\hat{\boldsymbol{x}}_0$ is the predicted $\boldsymbol{x}_0$ from $\boldsymbol{x}_t$ using Tweedie's formula[37]

$$\begin{aligned} \hat{\boldsymbol{x}}_0 &= \mathbb{E}[\boldsymbol{x}_0|\boldsymbol{x}_t] \\ &= \frac{1}{\sqrt{\bar{\alpha}_t}} \left[ \boldsymbol{x}_t + (1 - \bar{\alpha}_t)\boldsymbol{s}_{\theta,t}(\boldsymbol{x}_t) \right] \end{aligned} \tag{10}$$

Once the subproblem (Eq. 9) is solved, one can obtain a data-consistent sample $\widetilde{\boldsymbol{x}}_0$ in the target distribution $p(\boldsymbol{x}_0)$. This process can be seen as a projection operation from the estimated $\hat{\boldsymbol{x}}_0$ to the measurement-consistent set. Then we will continue the stochastic sampling of the reverse SDE from the updated estimation $\widetilde{\boldsymbol{x}}_0$ and previously generated transition $\boldsymbol{x}_t$ according to the VP-SDE. This strategy to enforce data consistency is similar to the unsupervised metal artifacts reduction for CT.[38]

In this way, we formulated unsupervised CBCT-based sCT generation in a Bayesian framework that incorporates the diffusion posterior sampling and the data consistency between sCT and CBCT. The score model $\boldsymbol{s}_{\theta,t}(\cdot)$ was trained on pCT images from the same patient, thus it is called patient-specific score-based prior. The training and sampling stages of the proposed method are summarized in Figure 1.

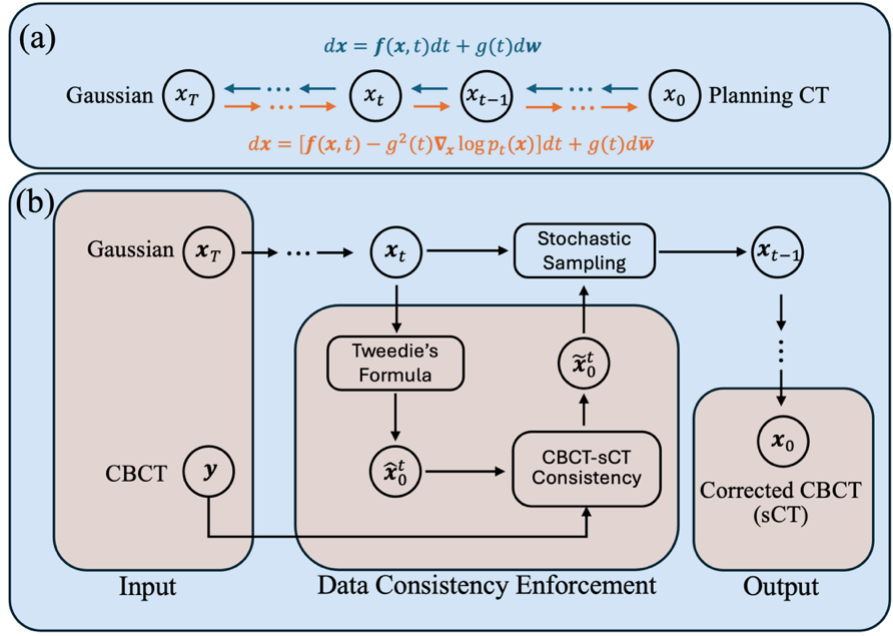

**Figure 1**. (a) Unconditional score matching of the patient-specific diffusion prior. (b) Generation of sCT through the incorporation of data consistency into the reverse SDE.

## 2.3 Volumetric sCT generation using the 2 dimensional (2D) score function

Due to the limited volumetric data availability and GPU memory limitation, it is preferred to train the patient-specific prior ($s_{\theta,t}(\cdot)$) using 2D images and perform Bayesian sCT generation in a slice-by-slice manner. However, there is an inherent issue if the generation of each slice is independent. The generated sCT volume will not be coherent between transverse slices due to the stochastic property of the reverse SDE, leading to artifacts in the coronal and sagittal views.

To resolve this issue, we aim to leverage the advantages of compressed sensing-based iterative reconstruction in the sCT generation.[39,40] Specifically, we introduce the total-variation (TV)-norm regularization of the sCT volume in the $z$-axis ($\|\cdot\|_{TV_z}$), to enforce sparsity on the finite difference between different slices. Mathematically speaking, the projection step in Eq. (9) is updated with an additional TV-norm minimization along the $z$-axis

$$\min_{\widetilde{x}_0} \phi' = \|\widetilde{x}_0 - y\|_2^2 + \eta\|\widetilde{x}_0 - \widehat{x}_0\|_2^2 + \lambda D_z(\widetilde{x}_0) \tag{11}$$

where $D_z$ denotes the finite difference matrix along the z-axis. The subproblem in Eq. (11) is solved by the fast iterative shrinkage-thresholding algorithm (FISTA).[41] The unsupervised training of the unconditional score model is summarized in Algorithm 1.1. The CBCT correction via diffusion model sampling is summarized in Algorithm 1.2.

**Algorithms 1**. Training and sampling stages of the proposed CBCT correction method.

| **Algorithm 1.1** Training | **Algorithm 1.2** Sampling (Correction) |
|---|---|
| 1: **repeat** | 1: $x_T \sim \mathcal{N}(0, \boldsymbol{I})$ |
| 2: $\quad x_0 \sim p(x_0)$ | 2: **for** $t = T, \cdots, 1$ **do** |
| 3: $\quad t \sim U([0,1])$ | 3: $\quad \hat{x}_0(x_t) = \frac{1}{\sqrt{\bar{\alpha}_t}}\left(x_t - \sqrt{1-\bar{\alpha}_t} \cdot \epsilon_{\theta,t}(x_t)\right)$   -Tweedie's formula |
| 4: $\quad \epsilon \sim \mathcal{N}(0, \boldsymbol{I})$ | 4: $\quad \tilde{x}_0(x_t) = \mathrm{argmin}_{\tilde{x}_0}\, \phi^{'}$     -Data consistency |
| 5: $\quad x_t = \sqrt{\bar{\alpha}_t}x_0 + \sqrt{1-\bar{\alpha}_t}\epsilon$ | 5: $\quad x_{t-1} \sim \mathcal{N}\left(x_{t-1}; \tilde{\mu}_t\big(x_t, \tilde{x}_0(x_t)\big), \tilde{\beta}_t \boldsymbol{I}\right)$    -Stochastic resampling |
| 6: $\quad$ Take an optimization step on | 6: **end for** |
| $\qquad \nabla_\theta \left\| \epsilon_{\theta,t}(x_t) - \epsilon \right\|^2$ | 7: **return** $x_0$ |
| 7: **until** converged | |

## 3. Evaluation

### 3.1 Data acquisition and preparation

In this retrospective study, 1600 CBCT and 1873 pCT slices were collected from 13 head and neck (H&N) cancer patients, 1000 CBCT and 1351 pCT slices were collected from 10 pancreatic cancer patients, and 1200 CBCT and 1643 pCT slices were collected from 12 lung cancer patients, all of whom were treated at Emory University Winship Cancer Institute. The selected lung cancer patients underwent SBRT using the breath-hold technique to minimize motion-induced artifacts. All CT scans were acquired on Siemens SOMATOM Definition AS with 120 kVp, and all CBCT scans were acquired using the on-board imager (OBI) system on Varian TrueBeam or Edge with 100 kVp. The pCT and CBCT were resampled to the same size of $1.0 \times 1.0 \times 1.0$ mm$^3$. For reference, pCT images were deformably registered to the corresponding CBCT images using Velocity AI 3.2.1.

### 3.2 Evaluation metrics

For quantitative evaluations, we calculated the MAE, PSNR, and NCC between the corrected CBCT and dpCT images, which were taken as the ground truth. The metrics are defined as follows:

$$MAE = \frac{1}{n_x n_y}\sum_{i,j}^{n_x, n_y}|sCT(i,j) - CT(i,j)| \tag{12}$$

$$PSNR = 10 \times \log_{10}\left(\frac{MAX^2}{\frac{1}{n_x n_y}\sum_{i,j}^{n_x,n_y}|sCT(i,j)-CT(i,j)|^2}\right) \tag{13}$$

$$NCC = \frac{1}{n_x n_y}\sum_{i,j}^{n_x,n_y}\frac{(sCT(i,j)-\overline{sCT})(CT(i,j)-\overline{CT})}{\sigma_{sCT}\sigma_{CT}} \tag{14}$$

where $sCT(i,j)$ and $CT(i,j)$ are the value of pixel $(i,j)$ in the sCT and CT respectively. $n_x n_y$ is the total number of pixels. $MAX$ is the maximum pixel value in the sCT and CT images. $\overline{sCT}$ and $\overline{CT}$ are the mean of sCT and CT images. $\sigma_{sCT}$ and $\sigma_{CT}$ are the standard deviation of sCT and CT images. MAE is the magnitude of the voxel-based Hounsfield unit (HU) difference between the original CT and the sCT. PSNR measures if the predicted sCT intensity is evenly or sparsely distributed. NCC is a measure of similarity between CT and sCT as a function of displacement.

### 3.3 Comparison studies

To verify the superiority of the proposed unsupervised CBCT correction method, we compared it with another two unsupervised learning schemes. The first one is iterative latent variable refinement (ILVR),[42] which is a popular technique for conditional image generation using unsupervised diffusion model. To demonstrate the efficacy of the patient-specific strategy, the second compared method is the proposed method without patient-specific image prior, i.e., the diffusion model is trained on pCTs from other patients. To verify the feasibility of the proposed strategy for 3D imaging using the 2D prior, the proposed method without the TV-norm minimization along the $z$-axis was implemented for comparison. The sampling stages of compared CBCT correction methods are summarized in Algorithms 2 and 3, where $\psi(\cdot)$ in the ILVR method denotes the linear low-pass filter consisting of sequent downsampling and upsampling operators. Of note, the latent variable refinement is also performed in the target CT distribution, i.e., $p(x_0)$.

**Algorithms 2 and 3**. Sampling stages of compared CBCT correction methods.

| **Algorithm 2** ILVR | **Algorithm 3** Proposed method w/o TVz |
|---|---|
| 1: $x_T \sim \mathcal{N}(0, \boldsymbol{I})$ | 1: $x_T \sim \mathcal{N}(0, \boldsymbol{I})$ |
| 2: **for** $t = T, \cdots, 1$ **do** | 2: **for** $t = T, \cdots, 1$ **do** |
| 3: $\quad \hat{x}_0(x_t) = \frac{1}{\sqrt{\bar{\alpha}_t}}\left(x_t - \sqrt{1-\bar{\alpha}_t} \cdot \epsilon_{\theta,t}(x_t)\right)$ | 3: $\quad \hat{x}_0(x_t) = \frac{1}{\sqrt{\bar{\alpha}_t}}\left(x_t - \sqrt{1-\bar{\alpha}_t} \cdot \epsilon_{\theta,t}(x_t)\right)$ |
| 4: $\quad \tilde{x}_0(x_t) = \psi(y) + (I-\psi)\hat{x}_0(x_t)$ | 4: $\quad \tilde{x}_0(x_t) = \text{argmin } \phi(\hat{x}_0, \boldsymbol{y})$ |
| 5: $\quad x_{t-1} \sim \mathcal{N}\left(x_{t-1}; \tilde{\mu}_t(x_t, \tilde{x}_0(x_t)), \tilde{\beta}_t \boldsymbol{I}\right)$ | 5: $\quad x_{t-1} \sim \mathcal{N}\left(x_{t-1}; \tilde{\mu}_t(x_t, \tilde{x}_0(x_t)), \tilde{\beta}_t \boldsymbol{I}\right)$ |
| 6: **end for** | 6: **end for** |
| 7: **return** $x_0$ | 7: **return** $x_0$ |

### 3.4 Implementation details

Instead of training $T$ totally different networks to predict $\epsilon_{\theta,t}(x_t, y)$ at each step, a single noise-prediction model with time-embedding was used in all $T$ steps. A U-net structure with attention modules and residual blocks was used to predict the noise in each time step.[43] For all score-based models, the total number of discrete time steps $T$ was set to 1000 and the noise variance was linearly scheduled from $\beta_1 = 10^{-4}$ to $\beta_T = 0.02$. The batch size was fixed at 2 and the drop-out ratio was set to 0.2. For the subproblem of Eq. (11), $\eta(t)$ is set to $(T-t)/T$ to gradually reduce the guidance of CBCT images during the sampling process, and $\lambda$ is set to a constant of $10^{-3}$ to balance the tradeoff between the data fidelity and TV-norm regularization. The linear low-pass filtering operator in the ILVR method was bicubic downsampling and upsampling function with a factor of 8. All the pCT and CBCT images were normalized to [-1, 1] before the model training and inference. Random shifts were performed on pCT images during the training stage. All the experiments were conducted using PyTorch 1.12 on an 80GB Nvidia A100 GPU.

### 4. Results

## 4.1 Visual quality improvements

Figure 2 demonstrates the performance of artifact reduction in the H&N patient study. Overall, the noise level was significantly suppressed by the proposed method. As indicated by the red arrows on CBCTs, streaking and shading artifacts were observed in original CBCT images and were efficiently reduced in the generated sCTs. For the third slice, the proposed method nearly corrected all the metal artifacts caused by the implants. As shown in the last slice, residual artifacts could be observed in the generated sCT for slices with server metal artifacts due to the limitation of image-domain correction. Compared to the generated sCT images, the dpCTs showed a mismatch from CBCTs due to the anatomical change between the pCT and CBCT scans.

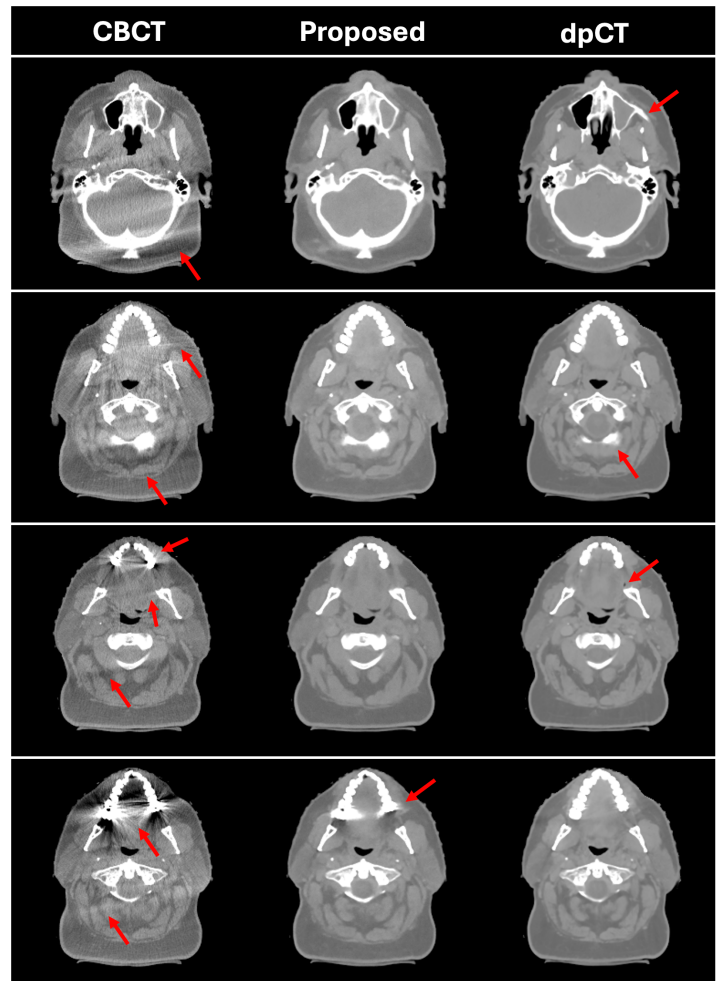

**Figure 2**. Selected results in the study of H&N patients. The left to right columns list the CBCT, generated sCT using the proposed method, and reference dpCT images. Red arrows indicate the artifacts or the structural mismatch. Display window: [-500 500] HU.

Selected slices from the pancreatic patient study are presented in Figure 3. For CBCTs, severe scattering and motion-induced artifacts were observed near the air pockets and blurred the structures. As shown in the second column ("Proposed"), the proposed method significantly corrected these artifacts and blurring while preserving the anatomical structure efficiently. For the dpCTs, obvious mismatches of both the body contour and the inner structure were observed, as indicated by the red arrows.

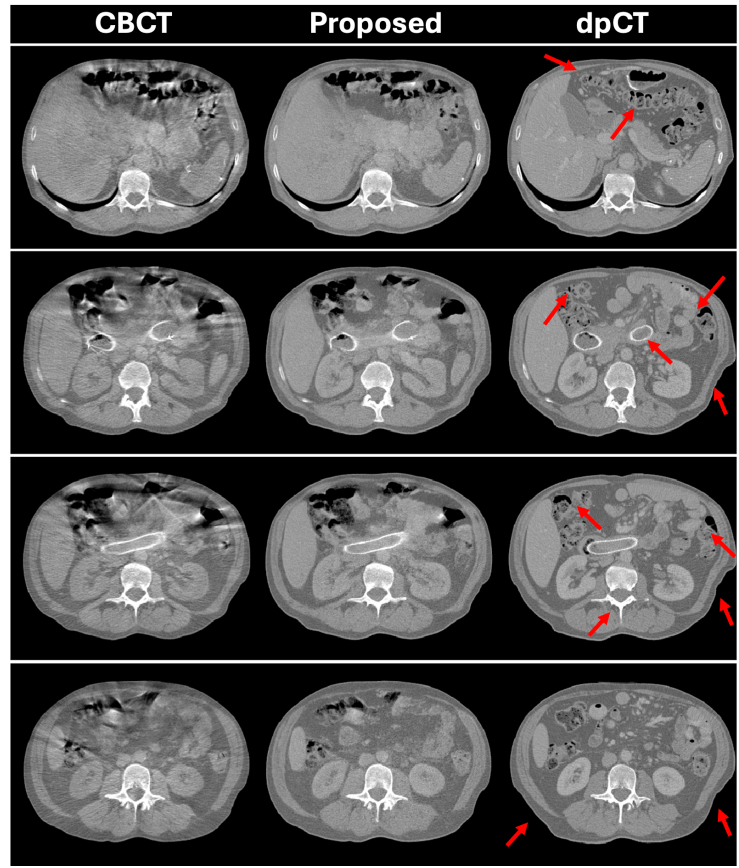

**Figure 3**. Selected slices in the study of pancreatic patients. Red arrows indicate the structural mismatch between dpCT and original CBCT. Display window: [-500 500] HU.

### 4.2 Quantitative analysis

In this work, the study of lung SBRT patients was used for the quantitative analysis. Selective slices were summarized in Figure 4. Consistent with the studies of H&N and pancreatic patients, streaking and shading artifacts in the CBCTs were efficiently reduced by the proposed method in the generated sCT images. For the dpCTs, mismatches of inner structures from the CBCTs were observed while the body contour remained the same, as indicated by the red arrows.

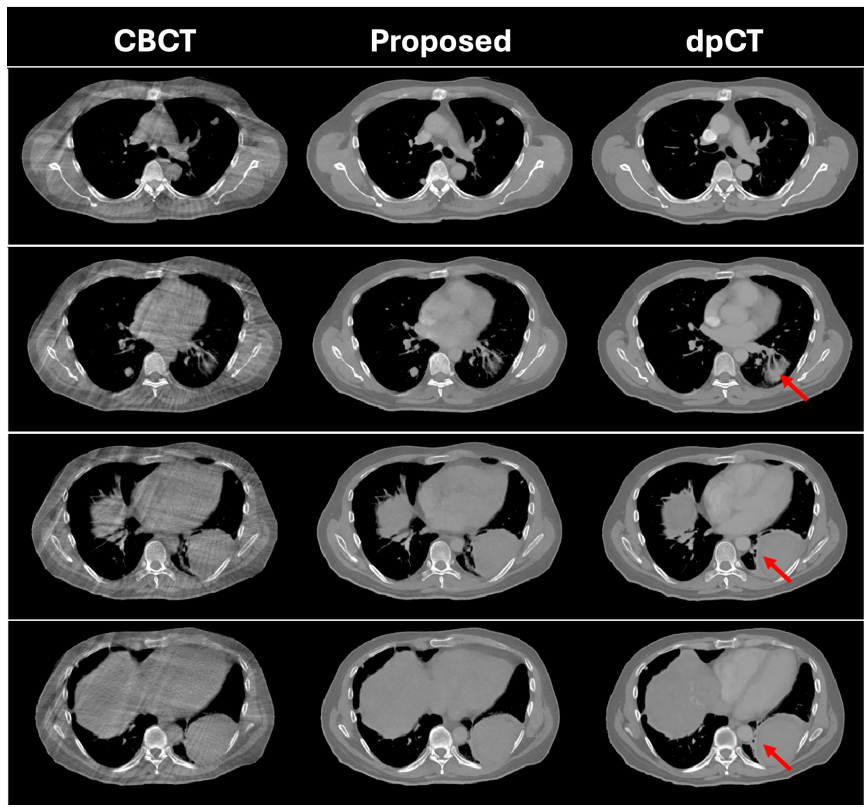

**Figure 4**. Selected results in the study of lung SBRT patients. Red arrows indicate the anatomical mismatches between dpCT and original CBCT. Display window: [-500 500] HU.

In the lung SBRT patient study, the HU accuracy was improved from MAE of 91 HU for the CBCT to 50 HU for the generated sCT. For the original CBCT, the PSNR and NCC were 28 dB and 0.98, which were improved to 31 dB and 0.99 for the corrected images. The results of quantitative metrics were tabulated in Table 1, showing the quantitative improvement of the generated sCT images compared to the original CBCTs.

**Table I.** Numerical comparison among CBCT and sCT generated by different methods.

|                          | MAE (HU)     | PSNR (dB)    | NCC       |
| ------------------------ | ------------ | ------------ | --------- |
| CBCT                     | 91.38±14.68  | 27.63±1.40   | 0.98±0.01 |
| ILVR                     | 99.01±15.10  | 25.17±1.39   | 0.95±0.02 |
| Proposed w/o $TV_z$      | 49.43±10.77  | 31.26±1.18   | 0.99±0.01 |
| Proposed                 | 50.45±10.96  | 31.16±1.21   | 0.99±0.01 |

## 4.3 Comparison studies

sCT images generated by different unsupervised diffusion models were summarized in Figure 5. For the results of ILVR scheme, the streaking artifacts were reduced but the structures were blurred in the transverse slices. Furthermore, the generated volume was incoherent between transverse slices for the stochastical generation of each slice, as shown in the coronal views. There was no significant difference between the transverse slices generated by the proposed method with and without TV regularization, as shown in the third and fourth columns in the top two rows. This is reasonable because the TV regularization operates on the orthogonal direction and has no effect inside the transverse plane. However, the superiority of the TV regularization was shown in the coronal views while the structures were coherent and consistent between different transverse slices, as shown in the bottom two rows.

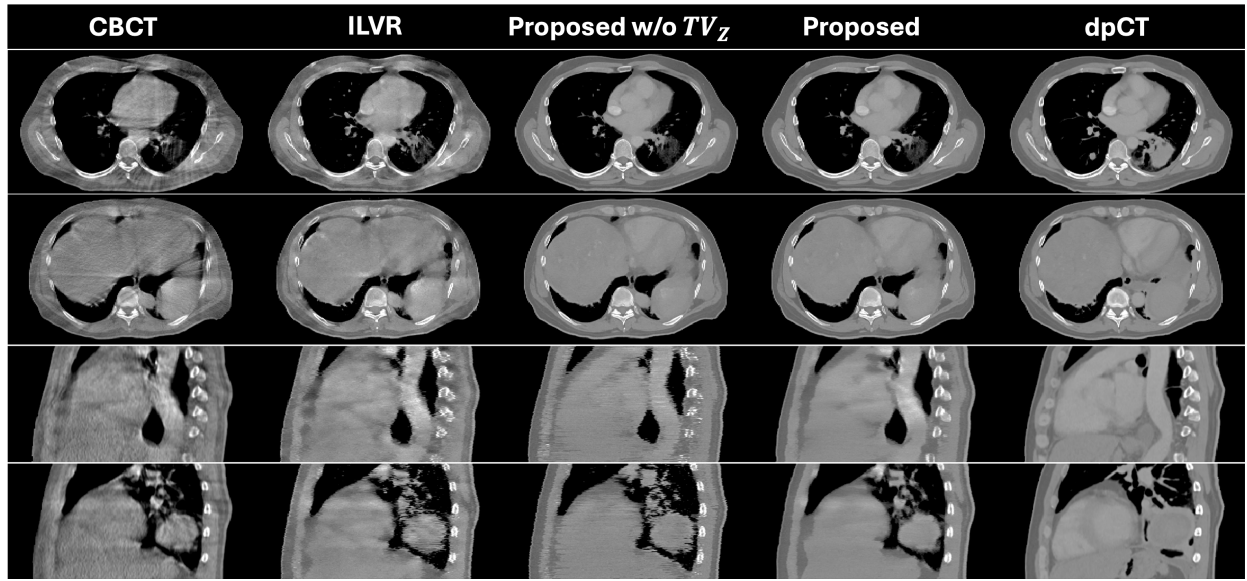

**Figure 5**. Summary of sCT generated using different score-based diffusion models in the study of lung SBRT patients. The top two rows show the axial slices, and the bottom two rows list the coronal views. Display windows: [-500 500] HU.

## 5. Discussions

This work proposed an unsupervised framework for CBCT correction in the image domain. A patient-specific score model was unconditionally trained using the same patient's pCT images and served as the image prior during the iterative CBCT correction process. A customized TV regularization along the $z$-axis was introduced to enforce the coherence between different transverse slices. To the best of our knowledge, this is the first work to introduce the patient-specific model into unsupervised CBCT correction.

The significance of this work is three-fold. First, we formulated the CBCT correction task as a prior-regularized inverse problem, making the data consistency explicitly enforced during the sCT generation. Second, we proposed an unsupervised Bayesian framework to solve the inverse problem based on the patient-specific score-based diffusion model, eliminating the image registration between the pCT and CBCT required by supervised learning. Third, we realized the volumetric correction in a slice-by-slice manner and suppressed the incoherence between transverse slices via a customized TV regularization, tackling the limitations of insufficient volumetric data and the requirement of large GPU memory for a whole-volume model.

Both the proposed method and previously reported conditional DDPM-based method try to simulate the conditional score function $\nabla_x \log p(x_t|y)$ via the deep learning model.[28] The previous work directly trained the conditional score model with paired CBCT-dpCT data via supervised learning. On the contrary, the proposed method first trained an unconditional score model via unsupervised learning and simulated the conditional score model by leveraging Bayes' rule.

As shown in the last row of Figure 2, residual metal artifacts can be observed in the sCT generated by the proposed method. This is a limitation of the proposed image-domain correction method. Specifically, the corruption of metal implants is localized in the projection data, but the artifacts are spread to surrounding areas because the Feldkamp-Davis-Kress (FDK) reconstruction is a non-local operator. In the image-domain correction, we limited the Euclidean distance between sCT and CBCT to enforce the data consistency. However, this will lead to residual artifacts if the original CBCT is severely corrupted, like the last slice in Figure 2.

Simultaneous CBCT reconstruction and correction is a solution to overcome this issue. This modification is straightforward by replacing the image consistency objective function in Eq. (9) with a projection consistency of $\|F\hat{x}_0 - p\|_2^2$, where $p$ is the projection data and $F$ is the forward projection operator during the CBCT scan. The following unsupervised framework to solve the inverse problem will be kept the same as it is in this image-domain correction work. Then we can discard the projections within the metal trace in the modified data-fidelity term and solve the ill-posed inverse problem by the patient-specific score-based image prior. In this way, we will avoid the severe metal artifacts in the original CBCT reconstructed by FDK algorithm and will reduce their impacts on the image correction via the image consistency of Eq. (9). Besides, this strategy can also be applied for other ill-posed CBCT reconstruction problems, such as the limited-angle, scout-view, and single-view CBCT imaging.[32,44-46] This will be the focus of our future research.

One of the common issues in the sCT generation works is that the mismatch between CBCT and dpCT images, leading to unfair results in the quantitative analysis when employing the dpCT as the ground truth.[47] In our following-up works, we will include Monte-Carlo simulation studies and physical phantom studies for the quantitative analysis, in which way we can provide more reasonable results for the exactly matched ground truth.

The time-consuming reverse sampling is a major bottleneck for the clinical application of score-based diffusion models. The acceleration of diffusion models is a hot topic in deep learning research and has been demonstrated in medical imaging applications.[48-50] For example, Xia *et al* achieved accelerated diffusion model for low-dose CT denoising using the fast ordinary differential equation (ODE) solver.[51] Pan et al achieved high-efficiency low-dose PET denoising using an accelerated diffusion model, consistency model.[26,52] We will incorporate these acceleration strategies into the CBCT reconstruction and correction framework for faster imaging. The other direction of our future research is to investigate the feasibility of different strategies for the enforcement of data consistency. This is an active research area to solve the inverse problem using unconditional score models. Besides the proposed score-based image prior for CBCT correction, we will also investigate the feasibility of dual-domain priors (projection domain and image domain) for simultaneous CBCT reconstruction and correction. Specifically, two score-based priors will be trained separately using the pCT images and cone-beam projections of the pCT volume. During the reconstruction process, both the projection and the image will be iteratively updated and corrected with the guidance of the patient-specific projection and image priors.

## 6. Conclusion

In this work, we designed an unsupervised framework for CBCT correction with a patient-specific score-based image prior and a customized TV regularization. The feasibility of the proposed method has been verified using image data from H&N, pancreatic, and lung cancer patients. The proposed method significantly reduced various artifacts and improved the HU accuracy of CBCT images, making it suitable for CBCT-based contouring and dose calculation. The proposed method has the potential to enable online dose verification and replanning in CBCT-based adaptive radiotherapy using existing CBCT hardware used on the linear accelerator (LINAC).

**Acknowledgments**

This research is supported in part by the National Institutes of Health under Award Number R01CA272991, R01EB032680, R37CA272755 and U54CA274513. K.Y. was supported by the RSNA Research Fellow Grant and ASTRO-LUNGevity Foundation Radiation Oncology Seed Grant.

## References

1. Haworth A, Paneghel A, Herschtal A, et al. Verification of target position in the post-prostatectomy cancer patient using cone beam CT. *Journal of medical imaging and radiation oncology*. 2009;53(2):212-220.

2. Schulze R, Heil U, Groβ D, et al. Artefacts in CBCT: a review. *Dentomaxillofacial Radiology*. 2011;40(5):265-273.

3. Cho PS, Johnson RH, Griffin TW. Cone-beam CT for radiotherapy applications. *Physics in Medicine & Biology*. 1995;40(11):1863.

4. Barrett JF, Keat N. Artifacts in CT: recognition and avoidance. *Radiographics*. 2004;24(6):1679-1691.

5. Wang H, Dong L, Lii MF, et al. Implementation and validation of a three-dimensional deformable registration algorithm for targeted prostate cancer radiotherapy. *International Journal of Radiation Oncology\* Biology\* Physics*. 2005;61(3):725-735.

6. Siewerdsen JH, Moseley D, Bakhtiar B, Richard S, Jaffray DA. The influence of antiscatter grids on soft-tissue detectability in cone-beam computed tomography with flat-panel detectors: Antiscatter grids in cone-beam CT. *Medical physics*. 2004;31(12):3506-3520.

7. Cai W, Ning R, Conover D. Scatter correction using beam stop array algorithm for cone-beam CT breast imaging. SPIE; 2006:1157-1165.

8. Zhu L, Xie Y, Wang J, Xing L. Scatter correction for cone-beam CT in radiation therapy. *Medical physics*. 2009;36(6Part1):2258-2268.

9. Xu Y, Bai T, Yan H, et al. A practical cone-beam CT scatter correction method with optimized Monte Carlo simulations for image-guided radiation therapy. *Physics in Medicine & Biology*. 2015;60(9):3567.

10. Nomura Y, Xu Q, Shirato H, Shimizu S, Xing L. Projection-domain scatter correction for cone beam computed tomography using a residual convolutional neural network. *Med Phys*. Jul 2019;46(7):3142-3155. doi:10.1002/mp.13583

11. Jiang Y, Zhang Y, Luo C, et al. A generalized image quality improvement strategy of cone-beam CT using multiple spectral CT labels in Pix2pix GAN. *Phys Med Biol*. May 17 2022;67(11)doi:10.1088/1361-6560/ac6bda

12. Yu L, Zhang Z, Li X, Xing L. Deep sinogram completion with image prior for metal artifact reduction in CT images. *IEEE Transactions on Medical Imaging*. 2020;40(1):228-238.

13. Gao L, Xie K, Sun J, et al. Streaking artifact reduction for CBCT-based synthetic CT generation in adaptive radiotherapy. *Med Phys*. Oct 2 2022;doi:10.1002/mp.16017

14. Harms J, Lei Y, Wang T, et al. Paired cycle-GAN-based image correction for quantitative cone-beam computed tomography. *Med Phys*. Sep 2019;46(9):3998-4009. doi:10.1002/mp.13656

15. Nomura Y, Xu Q, Peng H, et al. Modified fast adaptive scatter kernel superposition (mfASKS) correction and its dosimetric impact on CBCT-based proton therapy dose calculation. *Medical Physics*. 2020;47(1):190-200.

16. Harms J, Lei Y, Wang T, et al. Paired cycle-GAN-based image correction for quantitative cone-beam computed tomography. *Medical physics*. 2019;46(9):3998-4009.

17. Liang X, Chen L, Nguyen D, et al. Generating synthesized computed tomography (CT) from cone-beam computed tomography (CBCT) using CycleGAN for adaptive radiation therapy. *Physics in Medicine & Biology*. 2019;64(12):125002.

18. Liu Y, Lei Y, Wang T, et al. CBCT-based synthetic CT generation using deep-attention cycleGAN for pancreatic adaptive radiotherapy. *Medical physics*. 2020;47(6):2472-2483.

19. Chen L, Liang X, Shen C, Jiang S, Wang J. Synthetic CT generation from CBCT images via deep learning. *Medical physics*. 2020;47(3):1115-1125.

20. Zhang Y, Yue N, Su MY, et al. Improving CBCT quality to CT level using deep learning with generative adversarial network. *Medical physics*. 2021;48(6):2816-2826.

21. Chen L, Liang X, Shen C, Nguyen D, Jiang S, Wang J. Synthetic CT generation from CBCT images via unsupervised deep learning. *Physics in Medicine & Biology*. 2021;66(11):115019.

22. Kazerouni A, Aghdam EK, Heidari M, et al. Diffusion models in medical imaging: A comprehensive survey. *Medical Image Analysis*. 2023:102846.

23. Pan S, Wang T, Qiu RL, et al. 2D medical image synthesis using transformer-based denoising diffusion probabilistic model. *Physics in Medicine & Biology*. 2023;68(10):105004.

24. Pan S, Abouei E, Wynne J, et al. Synthetic CT generation from MRI using 3D transformer-based denoising diffusion model. *Medical Physics*. 2024;51(4):2538-2548.

25. Chang C-W, Peng J, Safari M, et al. High-resolution MRI synthesis using a data-driven framework with denoising diffusion probabilistic modeling. *Physics in Medicine & Biology*. 2024;69(4):045001.

26. Pan S, Abouei E, Peng J, et al. Full-dose whole-body PET synthesis from low-dose PET using high-efficiency denoising diffusion probabilistic model: PET consistency model. *Medical Physics*. 2024;

27. Ho J, Jain A, Abbeel P. Denoising diffusion probabilistic models. *Advances in neural information processing systems*. 2020;33:6840-6851.

28. Peng J, Qiu RL, Wynne JF, et al. CBCT-Based synthetic CT image generation using conditional denoising diffusion probabilistic model. *Medical physics*. 2024;51(3):1847-1859.

29. Song Y, Shen L, Xing L, Ermon S. Solving Inverse Problems in Medical Imaging with Score-Based Generative Models. 2021:

30. Chung H, Kim J, Mccann MT, Klasky ML, Ye JC. Diffusion posterior sampling for general noisy inverse problems. *arXiv preprint arXiv:220914687*. 2022;

31. Chung H, Lee S, Ye JC. DECOMPOSED DIFFUSION SAMPLER FOR ACCELERAT-ING LARGE-SCALE INVERSE PROBLEMS.

32. Xie J, Shao H-C, Li Y, Zhang Y. Prior Frequency Guided Diffusion Model for Limited Angle (LA)-CBCT Reconstruction. *arXiv preprint arXiv:240401448*. 2024;

33. Song B, Kwon SM, Zhang Z, Hu X, Qu Q, Shen L. Solving inverse problems with latent diffusion models via hard data consistency. *arXiv preprint arXiv:230708123*. 2023;

34. Chung H, Sim B, Ryu D, Ye JC. Improving diffusion models for inverse problems using manifold constraints. *Advances in Neural Information Processing Systems*. 2022;35:25683-25696.

35. Vincent P. A connection between score matching and denoising autoencoders. *Neural computation*. 2011;23(7):1661-1674.

36. Song Y, Sohl-Dickstein J, Kingma DP, Kumar A, Ermon S, Poole B. Score-based generative modeling through stochastic differential equations. *arXiv preprint arXiv:201113456*. 2020;

37. Efron B. Tweedie's formula and selection bias. *Journal of the American Statistical Association*. 2011;106(496):1602-1614.

38. Liu X, Xie Y, Diao S, Tan S, Liang X. Unsupervised CT Metal Artifact Reduction by Plugging Diffusion Priors in Dual Domains. *IEEE Transactions on Medical Imaging*. 2024;

39. Sidky EY, Pan X. Image reconstruction in circular cone-beam computed tomography by constrained, total-variation minimization. *Physics in Medicine & Biology*. 2008;53(17):4777.

40. Chen Z, Jin X, Li L, Wang G. A limited-angle CT reconstruction method based on anisotropic TV minimization. *Physics in Medicine & Biology*. 2013;58(7):2119.

41. Beck A, Teboulle M. A fast iterative shrinkage-thresholding algorithm for linear inverse problems. *SIAM journal on imaging sciences*. 2009;2(1):183-202.

42. Choi J, Kim S, Jeong Y, Gwon Y, Yoon S. Ilvr: Conditioning method for denoising diffusion probabilistic models. *arXiv preprint arXiv:210802938*. 2021;

43. Dhariwal P, Nichol A. Diffusion models beat gans on image synthesis. *Advances in neural information processing systems*. 2021;34:8780-8794.

44. Montoya JC, Zhang C, Li Y, Li K, Chen GH. Reconstruction of three-dimensional tomographic patient models for radiation dose modulation in CT from two scout views using deep learning. *Medical physics*. 2022;49(2):901-916.

45. Shen L, Zhao W, Xing L. Patient-specific reconstruction of volumetric computed tomography images from a single projection view via deep learning. *Nature biomedical engineering*. 2019;3(11):880-888.

46. Pan S, Lo S-Y, Chang C-W, et al. Patient-specific 3D volumetric CBCT image reconstruction with single x-ray projection using denoising diffusion probabilistic model. SPIE; 2024:136-143.

47. Rusanov B, Hassan GM, Reynolds M, et al. Deep learning methods for enhancing cone-beam CT image quality toward adaptive radiation therapy: A systematic review. *Medical Physics*. 2022;49(9):6019-6054.

48. Song J, Meng C, Ermon S. Denoising diffusion implicit models. *arXiv preprint arXiv:201002502*. 2020;

49. Salimans T, Ho J. Progressive distillation for fast sampling of diffusion models. *arXiv preprint arXiv:220200512*. 2022;

50. Lu C, Zhou Y, Bao F, Chen J, Li C, Zhu J. Dpm-solver: A fast ode solver for diffusion probabilistic model sampling in around 10 steps. *Advances in Neural Information Processing Systems*. 2022;35:5775-5787.

51. Xia W, Lyu Q, Wang G. Low-Dose CT Using Denoising Diffusion Probabilistic Model for 20$\times$ Speedup. *arXiv preprint arXiv:220915136*. 2022;

52. Song Y, Dhariwal P, Chen M, Sutskever I. Consistency models. *arXiv preprint arXiv:230301469*. 2023;